\documentclass[twocolumn,superscriptaddress,nofootinbib,aps,prl,floatfix,preprintnumbers,amsmath,amssymb,groupedaddress]{revtex4}

\usepackage{dsfont}
\usepackage{epsfig}
\usepackage{slashed}
\usepackage{bbold}
\usepackage{psfrag}
\usepackage{color}
\PassOptionsToPackage{caption=false}{subfig}
\usepackage{subfig}
\usepackage{multirow}
\usepackage{booktabs}
\usepackage{hyperref}
\begin{document}

\def\as{\alpha_S}
\def\t{{\bar t}}
\def\Mtt{m_{t\bar t}}
\def\ytt{y_{t\bar t}}
\def\PT{p_{\rm T}}
\def\PTt{p_{{\rm T},t}}
\def\GeV{\, \rm GeV}

\title{High-Precision Differential Predictions for Top-Quark Pairs at the LHC}

\author{Michal Czakon}
\affiliation{Institut f\"ur Theoretische Teilchenphysik und Kosmologie,
RWTH Aachen University, D-52056 Aachen, Germany}

\author{David Heymes}
\affiliation{Cavendish Laboratory, University of Cambridge, Cambridge CB3 0HE, UK}

\author{Alexander Mitov}
\affiliation{Cavendish Laboratory, University of Cambridge, Cambridge CB3 0HE, UK}

\preprint{Cavendish-HEP-15/10, TTK-15-34}

\begin{abstract}
We present the first complete next-to-next-to-leading order (NNLO) QCD predictions for differential distributions in the top-quark pair production process at the LHC. Our results are derived from a fully differential partonic Monte Carlo calculation with stable top quarks which involves no approximations beyond the fixed-order truncation of the perturbation series. The NNLO corrections improve the agreement between existing LHC measurements [V. Khachatryan {\it et al.} (CMS Collaboration), Eur. Phys. J. C {\bf 75}, 542 (2015)] and standard model predictions for the top-quark transverse momentum distribution, thus helping alleviate one long-standing discrepancy. The shape of the top-quark pair invariant mass distribution turns out to be stable with respect to radiative corrections beyond NLO which increases the value of this observable as a place to search for physics beyond the standard model. The results presented here provide essential input for parton distribution function fits, implementation of higher-order effects in Monte Carlo generators as well as top-quark mass and strong coupling determination.
\end{abstract}
\maketitle

\section{Introduction}

There is remarkable overall agreement between standard model (SM) predictions for top-quark pair production and LHC measurements. Measurements of the total inclusive cross section at 7, 8, and 13 TeV \cite{Chatrchyan:2012bra,Chatrchyan:2013faa,Aad:2014kva,Khachatryan:2015uqb,ATLAS13TeV-sigmatot} agree well with next-to-next-to leading order (NNLO) QCD predictions \cite{Baernreuther:2012ws,Czakon:2012zr,Czakon:2012pz,Czakon:2013goa,Cacciari:2011hy,Czakon:2011xx}. Differential measurements of final state leptons and jets are generally well described by existing NLO QCD Monte Carlo (MC) generators. Concerning top-quark differential distributions, the description of the top-quark $\PT$ has long been in tension with data \cite{Chatrchyan:2012saa,Aad:2014zka,Aad:2015eia}; see also the latest differential measurements in the bulk \cite{Khachatryan:2015oqa} and  boosted top \cite{Aad:2015hna} regions. The first 13 TeV measurements have just appeared \cite{CMS13TeV-lep+jet,CMS13TeV-dilep} and they show similar results; i.e., MC predictions tend to be harder than data.

This ``$\PT$ discrepancy" has long been a reason for concern. Since the top quark is not measured directly, but is inferred from its decay products, any discrepancy between top-quark-level data and SM prediction implies that, potentially, the MC generators used in unfolding the data may not be accurate enough in their description of top-quark processes. With the top quark being a main background in most searches for physics beyond the SM (BSM), any discrepancy in the SM top-quark description may potentially affect a broad class of processes at the LHC, including BSM searches and Higgs physics. 

The main ``suspects" contributing to such a discrepancy are  higher order SM corrections to top-quark pair production and possible deficiencies in MC event generators. A goal of this work is to derive the NNLO QCD corrections to the top-quark $\PT$ spectrum at the LHC and establish if these corrections bridge the gap between LHC measurements, propagated back to top-quark level with {\it current} MC event generators, and SM predictions at the level of stable top quarks. 

\begin{figure}[h]
\centering
\hspace{0mm} 
\includegraphics[width=0.49\textwidth]{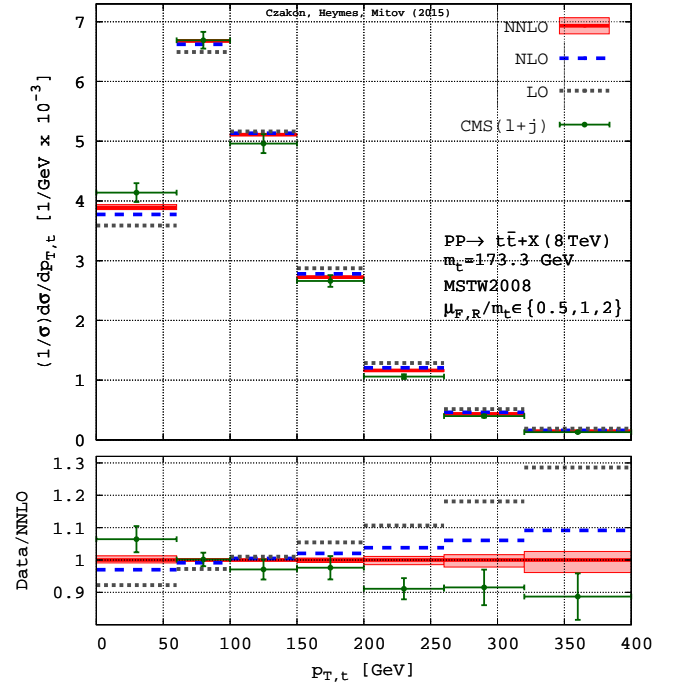} 
\caption{Normalized top-antitop $\PT$ distribution vs CMS lepton+jets data \cite{Khachatryan:2015oqa}. NNLO error band from scale variation only. The lower panel shows the ratios LO/NNLO, NLO/NNLO, and data/NNLO.}
\label{fig:PTt-norm}
\end{figure}
\begin{figure}[h]
\centering
\hspace{0mm} 
\includegraphics[width=0.49\textwidth]{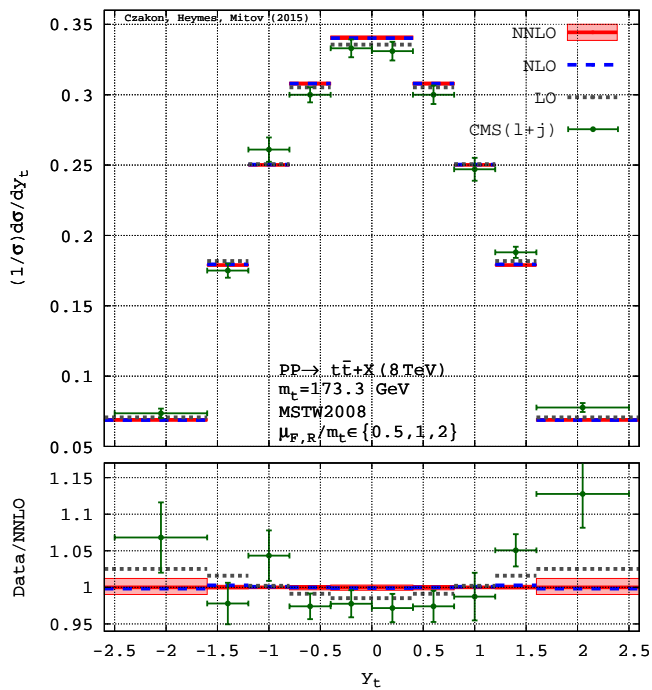} 
\caption{As in Fig.~\ref{fig:PTt-norm} but for the top-antitop rapidity.}
\label{fig:yt-norm}
\end{figure}
Our calculations are for the LHC at 8 TeV. They show that the NNLO QCD corrections to the top-quark $\PT$ spectrum are significant and must be taken into account for proper modeling of this observable. The effect of NNLO QCD correction is to soften the spectrum and bring it closer to the 8 TeV CMS data \cite{Khachatryan:2015oqa}. In addition to the top-quark $\PT$, all major top-quark pair differential distributions are studied as well.

\section{Details of the calculation}\label{sec:details}

In the context of our previous work on the top-quark forward-backward asymmetry at the Tevatron \cite{Czakon:2014xsa}, we have already preformed a complete differential calculation of NNLO QCD corrections to on-shell top-quark pair production. Unfortunately, our Tevatron setup turned out not to be sufficiently powerful to deal with the increased demands of the LHC configuration. One reason is that the cross-section is now dominated by gluon fusion instead of quark annihilation. The main cause lies, however, in the substantially higher collider energy, which raises the fraction of events with top quarks far away from threshold. For the latter, phase space integrals yield large logarithms of the ratio of the top-quark mass and the partonic center-of-mass energy. In consequence, the convergence rate of the numerical Monte Carlo integration is severely diminished.

The results presented in this Letter are obtained using a fresh complete implementation of the sector-improved residue subtraction scheme, \textsc{Stripper} \cite{Czakon:2010td,Czakon:2011ve}, in its four-dimensional formulation as developed in Ref.~\cite{Czakon:2014oma}. We note that the subtraction scheme relies on the known soft and collinear limits of tree-level and one-loop matrix elements \cite{Catani:1999ss,Campbell:1997hg,Catani:1998nv,DelDuca:1999ha,Czakon:2011ve,Bern:1994zx,Bern:1998sc,Kosower:1999xi,Kosower:1999rx,Bern:1999ry,Somogyi:2006db,Catani:2000pi,Catani:2000ef,Bierenbaum:2011gg}. It also exploits the singularity structure of one- and two-loop virtual amplitudes \cite{Ferroglia:2009ii}. Its main strength consists in preserving process independence and generality without requiring intricate analytic integration. The price of the obvious advantage is a numerical (as opposed to analytical) cancellation of the poles in the dimensional regularization parameter.

The process specific matrix elements for top-quark pair production in the Born approximation were obtained using the software from Ref.~\cite{Bury:2015dla}. We evaluated the four-point one-loop amplitudes ourselves, although they can also be found in Refs.~\cite{Korner:2004rr,Anastasiou:2008vd,Kniehl:2008fd}. The five-point one-loop amplitudes, on the other hand, were computed with a code used in the calculation of $pp \to t\t j$ at NLO \cite{Dittmaier:2007wz,Dittmaier:2008uj}. Finally, the two-loop matrix elements were taken in the form of numerical values on a dense grid supplemented with threshold and high-energy expansions from Refs.~\cite{Czakon:2008zk,Baernreuther:2013caa}. Notice that some partial analytical results are also known at two loops \cite{Bonciani:2013ywa,Bonciani:2010mn,Bonciani:2009nb,Bonciani:2008az}.

As for our setup, we use the top-quark pole mass $m_t=173.3\GeV$, the MSTW2008 parton distribution function (PDF) set \cite{Martin:2009iq}, and kinematics-independent scales with the central value $\mu_R=\mu_F=m_t$. The theoretical uncertainty is estimated with independent scale variation $\mu_R\neq\mu_F$ subject to the additional restriction $0.5<\mu_R/\mu_F<2$ \cite{Cacciari:2008zb}. The PDF uncertainty is not included. The above choice of scales, PDF set, and parameters is dictated mainly by reasons of backward compatibility with our previous work and the need for extensive checks at the level of intermediate and final results. In the future, we intend to consider various choices of running scales, PDF sets and errors as well as values of $m_t$.

We have checked that our calculation reproduces $\sigma_{\rm tot}$ from Refs.~\cite{Czakon:2013goa,Czakon:2012pz,Czakon:2012zr,Baernreuther:2012ws} for each value of $\mu_R,\mu_F$ with a precision around two per mil for the $\mathcal{O}(\alpha_s^4)$ contribution, which translates to about $2 \times 10^{-4}$ for the complete result. We have also verified the cancellation of infrared singularities in each histogram bin. At NLO, our calculation has been cross-checked with the MC generator MCFM \cite{Campbell:2012uf,Nason:1987xz}. The predicted NNLO $p_{T,t\bar t}$  distribution for nonvanishing transverse momentum is consistent with results for the NLO QCD corrections to $pp \to t\t j$ from Refs.~\cite{Melnikov:2010iu,Melnikov:2011qx,Hoeche:2013mua} and agrees with an independent evaluation using \textsc{Helac-Nlo} \cite{Bevilacqua:2011xh}. The new software also reproduces our previous Tevatron results.

\section{Results}

In the following we discuss the $\PTt, y_t, \Mtt$, and $\ytt$ differential distributions. We do not present the transverse momentum distribution of the top-quark pair since it can be obtained with readily available NLO tools applied to the $tt+j$ process. The $\PTt$ and $y_t$ distributions are assumed to be insensitive to the charge of the heavy quark; i.e., they are an average of the respective top- and antitop-quark distributions.

In Fig.~\ref{fig:PTt-norm} we show the prediction for the normalized $\PTt$ distribution computed in LO, NLO, and NNLO QCD, and compared to the most recent CMS data \cite{Khachatryan:2015oqa}. The corresponding top-quark rapidity distribution is shown in Fig.~\ref{fig:yt-norm}. As explained in the previous section, PDF variation has not been included in these results (or in any other results shown in this Letter). For clarity, in Figs.~\ref{fig:PTt-norm} and \ref{fig:yt-norm} the scale variation is only shown for the NNLO correction. When computing various perturbative orders we always use PDFs of matching order.

No overflow events are included in any of the bins shown in this Letter. The normalizations of the distributions in Figs.~\ref{fig:PTt-norm} and \ref{fig:yt-norm} are derived in such a way that the integral over the bins shown in these figures yields unity. Because of a slight difference in the bins, we note a small mismatch with respect to the measurements we compare to: for the top-quark $\PT$ distribution CMS has one additional bin $400\GeV < \PT < 500\GeV$ (not shown in Fig.~\ref{fig:PTt-norm}). This bin contributes only around 4 per mil to the normalization of the data and we neglect it in the comparison. The $y_t$ distribution computed by us extends to $|y_t|<2.6$. This last bin differs slightly from the corresponding CMS bin which extends to $|y_t|<2.5$. This mismatch is shown explicitly in Fig.~\ref{fig:yt-norm}.

\begin{figure}[t]
\centering
\hspace{0mm} 
\includegraphics[width=0.49\textwidth]{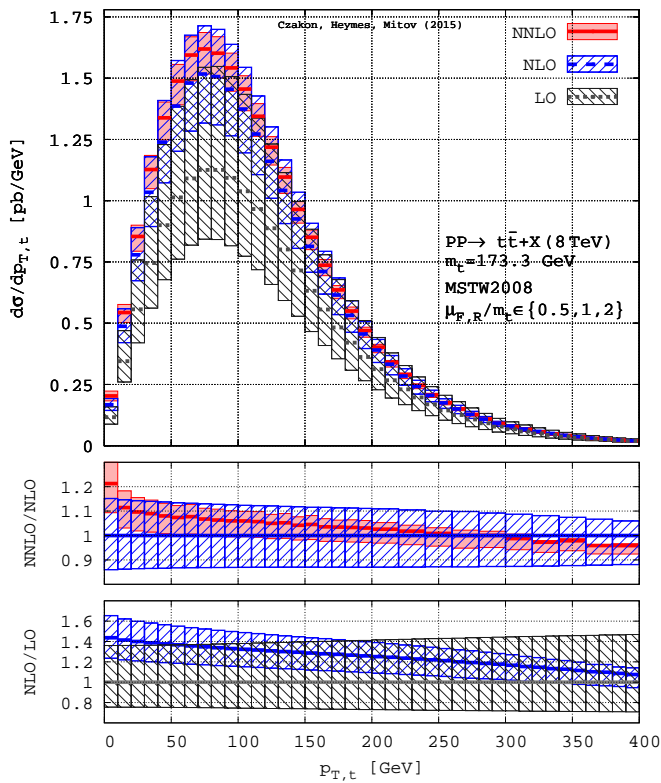} 
\caption{Top-antitop $\PT$ distribution in LO, NLO and NNLO QCD. Error bands from scale variation only.}
\label{fig:PTt}
\end{figure}
We observe that the inclusion of NNLO QCD corrections in the $\PTt$ distribution brings SM predictions closer to CMS data in all bins. In fact the two agree within errors in all bins but one (recall that the PDF error has not been included in Fig.~\ref{fig:yt-norm}). The case of the $y_t$ distribution is more intriguing; we observe in Fig.~\ref{fig:yt-norm} that the NNLO and NLO central values are essentially identical in the whole rapidity range (this is partly related to the size of the bins). Given the size of the data error, it does not appear that there is any notable tension between NNLO QCD and data. The apparent stability of this distribution with respect to NNLO radiative corrections will clearly make comparisons with future high-precision data very interesting.

\begin{figure}[t]
\centering
\hspace{0mm} 
\includegraphics[width=0.48\textwidth]{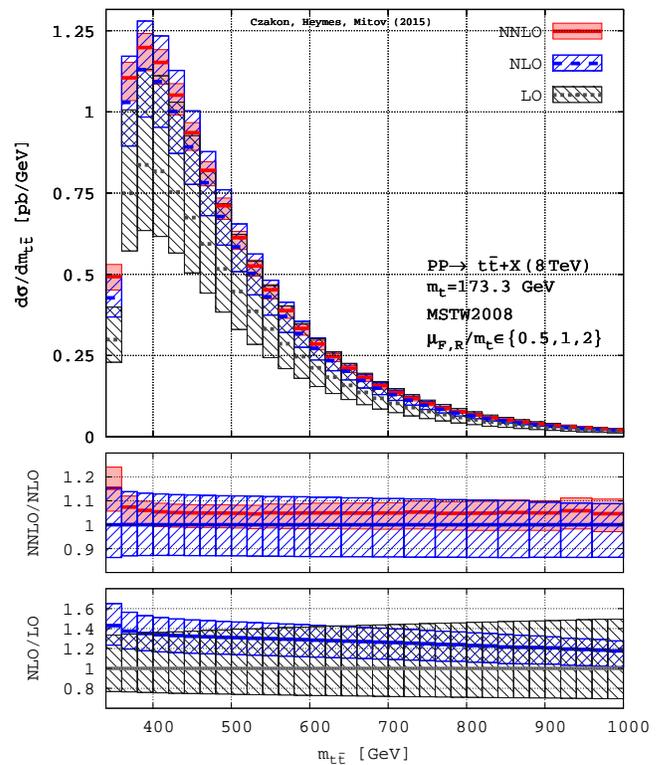} 
\caption{As in Fig.~\ref{fig:PTt} but for the top pair invariant mass.}
\label{fig:Mtt}
\end{figure}

\begin{figure}[t]
\centering
\hspace{0mm} 
\includegraphics[width=0.49\textwidth]{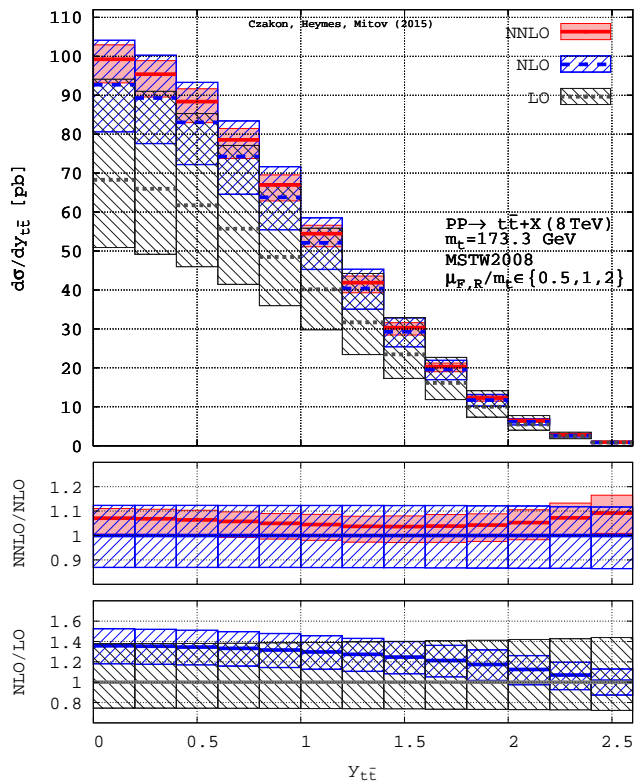} 
\caption{As in Fig.~\ref{fig:PTt} but for the top pair rapidity.}
\label{fig:ytt}
\end{figure}

We do not compare with the CMS data for the $\Mtt$ and $\ytt$ distributions since the mismatch in binning is more significant. Instead, in Figs.~\ref{fig:Mtt} and \ref{fig:ytt} we present the NNLO predictions for the absolute normalizations of these distributions. We stress that the bin sizes we present are significantly smaller than the ones in the existing experimental publications. This should make it possible to use our results in a variety of future experimental and theoretical analyses. For this reason, in Fig.~\ref{fig:PTt} we also present the absolute prediction for the top-quark $\PT$ distribution with much finer binning compared to the one in Fig.~\ref{fig:PTt-norm}. 

In Figs.~\ref{fig:PTt},\ref{fig:Mtt}, and \ref{fig:ytt} we show the scale variation for each computed perturbative order, together with the NLO and NNLO $K$ factors. In all cases one observes a consistent reduction in scale variation with successive perturbative orders. Importantly, we also conclude that our scale variation procedure is reliable, since NNLO QCD corrections are typically contained within the NLO error bands (and to a lesser degree for NLO with respect to LO). We also notice that the NNLO corrections do not affect the shape of the $\Mtt$ distribution. The stability of this distribution with respect to higher-order corrections makes it, among others, an ideal place to search for BSM physics. It will be very interesting to check if this property is maintained with dynamic scales and if it extends to higher $\Mtt$.

The $K$ factors in Figs.~\ref{fig:PTt} and \ref{fig:Mtt} show a peculiar rise at low $\PTt$ and $\Mtt$, respectively, which is due to soft gluon and Coulomb threshold effects. We do not investigate them in detail in the present work; related past studies include Refs.~\cite{Berger:1996ad,Kidonakis:1997gm,Laenen:1998qw,Kidonakis:2001nj,Ahrens:2010zv,Ahrens:2011mw,Kidonakis:2012rm,Kidonakis:2014pja,Hagiwara:2008df,Kiyo:2008bv}.

A feature of our calculation that needs to be addressed more extensively is the fact that we use fixed scales. Running scales are usually thought to be more appropriate for such a differential calculation. However, in this first work on the subject, we opt for the simplicity of fixed scales in order to perform checks with existing NNLO calculations. We intend to extend our result to dynamical scales, which typically involve the top transverse mass $\sqrt{p_T^2+m_t^2}$ and thus start to deviate from fixed scales at large $p_T$, in future publications. The result presented here, however, should not be affected substantially by such a change due to the limited kinematical range considered (for instance $\PTt<400\GeV$).

\section{Conclusions} 

In this Letter, we present for the first time NNLO accurate differential distributions for top-quark pair production at the LHC at 8 TeV. It is easy to conclude from the shown $K$ factors that our calculation is of very high quality (i.e. MC errors are small). Our result is exact in the sense that it {\it fully} includes all partonic channels contributing to NNLO and, moreover, includes them completely (in particular, we do not resort to the leading color approximation).

Partial NNLO results have been computed by two groups \cite{Abelof:2015lna,Abelof:2014jna,Bonciani:2015sha}. At the level of the total inclusive cross section these results agree with our previous calculations \cite{Baernreuther:2012ws,Czakon:2012zr,Czakon:2012pz,Czakon:2013goa}. Although highly desirable, a comparison at the differential level is not possible at present since in our current calculation we do not separate subsets of partonic reactions or implement the leading colour approximation. Additionally, various NNLO approximations exist in the literature \cite{Ahrens:2010zv,Ahrens:2011mw,Kidonakis:2012rm,Kidonakis:2014pja,Guzzi:2014wia,Broggio:2014yca}. A dedicated comparison with these approximate results would be valuable.

The results derived in this Letter would allow one to undertake a number of high-caliber phenomenological LHC analyses. Some examples are: validation of different implementations of higher-order effects in MC event generators, extraction of NNLO PDFs from LHC data, improved determination of the top-quark mass, and direct measurement of the running of $\alpha_S$ at high scales. Moreover, SM predictions with improved precision will enable a higher level of scrutiny of the SM with the help of LHC data as well as novel searches for BSM physics, possibly along the lines of Refs.~\cite{Czakon:2014fka,Aad:2014kva}. Finally, this result will serve as the basis for future inclusion of top-quark decay \cite{Gao:2012ja,Brucherseifer:2013iv}.

\acknowledgments\vskip4mm
We thank Stefan Dittmaier for kindly providing us with his code for the evaluation of the one-loop virtual corrections. M.C. thanks Emmanuel College Cambridge for hospitality during the completion of this work. A.M. thanks Durham University for hospitality during the completion of this work.
The work of M.C. was supported in part by grants of the DFG and BMBF. The work of D.H. and A.M. is supported by the UK Science and Technology Facilities Council [Grants No. ST/L002760/1 and No. ST/K004883/1].

\end{document}